\documentclass[onecolumn,draftclsnofoot,12pt]{IEEEtran}
\ifCLASSINFOpdf

\else

\fi

\usepackage{amsfonts,epsfig,amssymb}

\usepackage{caption}

\usepackage{graphicx}
\usepackage{epstopdf}

\usepackage{mathtools,bm,color}
\usepackage[numbers,sort&compress]{natbib}

\usepackage{algorithm}
\usepackage[noend]{algpseudocode}

\begin{document}

\title{Sparse Impulse Noise Minimization in UWB Communication using  Signal Characteristics }

\author{\IEEEauthorblockN{Sanjeev Sharma$^{1}$, Anubha Gupta$^{2}$ (Senior Member, IEEE)}  and Vimal Bhatia$^{1}$ (Senior Member, IEEE)\\
\IEEEauthorblockA{$^{1}$ \small{Indian Institute of Technology Indore, India 453552
\{phd1501102011,vbhatia\}@iiti.ac.in}
\and \\
\IEEEauthorblockA{$^{2}$ Indraprastha Institute of Information Technology-Delhi,
New Delhi-110020,
anubha@iiitd.ac.in}}}

\maketitle

\begin{abstract}
In many applications, ultra-wide band (UWB) system experiences impulse noise due to  surrounding physical noise sources. Therefore, a conventional receiver (correlator or matched filter) designed for additive Gaussian noise system  is not optimum for an impulse noise affected communication channel. In this paper, we propose a new robust receiver design that utilizes the received UWB signal cluster sparsity to mitigate impulse noise. Further, multipath channel diversity enhances the signal-to-noise ratio,  as compared to the single path after impulse noise removal in the proposed receiver design.
The proposed receiver is analyzed in time hopping binary phase shift keying UWB system and is compared with popular blanking non-linearity based receiver in
Bernoulli-Gaussian  impulse noise over
both single and multipath IEEE 802.15.4a channels.
Unlike existing designs, the
proposed receiver does not require any training sequence. The proposed receiver is observed to be robust with improved bit error rate performance as compared to a blanking receiver in the presence of impulse noise.
\end{abstract}
%
%

\IEEEpeerreviewmaketitle

\section{Introduction}

Performance of conventional receivers (correlator or matched filter), designed for additive white Gaussian noise (AWGN) channels, deteriorates in  harsh  environment such as industrial and mining due to impulse nature of noise plus interference
\cite{ding2013first, cheffena2016}.
Impulse noise in wireless communication systems like  WSNs (wireless sensor networks), IoT (Internet of Things), and M2M (machine-to-machine) deployed   in mining, industrial, home, power line, and underwater, occur due to ignition, lightening, hardware impairment, ice cracking etc.
Therefore, in the presence of impulse noise, performance of conventional receiver  deteriorates.
Hence,  in an impulse noise environment, some robust  signal pre-processing techniques are required
to ensure proper functionality, quality, and performance throughout the system's operation.

The degrading impact of impulse noise can be minimized or mitigated using non-linear techniques based mitigators. Non-linear impulse noise mitigation methods such as clipping and blanking  are simple. However, these methods are suboptimal and sensitive to the choice of threshold value
\cite{zhidkov2006, kim2006comparative,guney}. Some methods involve impulse noise estimation followed by subtraction from the received signal using null carriers or training data \cite{lin2013}.
However, the occurrence of impulse noise samples is completely  random. Hence, training or estimation based impulse noise mitigation  methods may not be useful in a practical system. In \cite{el2010, niranjayan2013,ekrem2007ultra}, various non-linear receiver structures are analyzed for impulse noise scenarios in UWB systems. However, their performance depends on the receiver model parameters accuracy and feasibility estimation.  In our earlier work \cite{san2016, sharma2017}, the sparsity of ultra-wide band (UWB) signal is exploited to mitigate impulse noise and narrowband interference effects in UWB systems.
However, the computational complexity for such receivers is large for a UWB signal vector with higher sampling frequency and frame duration.

In this paper, we  propose a  novel signal cluster-detection based  receiver design to mitigate impulse noise in a UWB system.
The received UWB signal forms clusters  due to signal propagation characteristics \cite{molish2006, kim2006comparative, san2016,ncc2017, yang2016variance,silva2016} and hence is also called as cluster sparse signal.
The proposed cluster detection algorithm easily differentiates between UWB signal cluster and  impulse noise.
The time-hopping binary phase shift keying (TH-BPSK) UWB system is considered for bit error rate (BER) performance analysis in the presence of Bernoulli-Gaussian (BG) impulse noise in
 AWGN and multipath IEEE 802.15.4a channels to validate the proposed algorithm.\\
\textit{Notations}: Small and bold small letters represent
a scalar and vector respectively.
$\left\Vert (\cdot) \right \Vert_{2}$ is the Euclidian norm of a signal $(\cdot)$, and $\mathcal{N}(\eta, \sigma^{2})$ represents the Gaussian probability distribution function (pdf) with mean $\eta$ and variance $\sigma^{2}$. Symbols $\langle .,.\rangle $ and ``$\ast$" represent the inner product and convolution between two vectors, respectively.
$Pr\{\mathcal{B}\}$ and $|(\cdot)|$ denote the probability of event $\mathcal{B}$
and the absolute magnitude value of $(\cdot)$, respectively.

\vspace{-1em}
\section{System Model}\label{sect2}
In this section, BG impulse noise and basic TH-BPSK UWB system models  are described. The BG  impulse noise $i(t)$ is  represented as \cite{san2016}
\begin{equation}\label{eq1}
i(t)=b(t)k(t),
\end{equation}
where $b(t)$ is the Bernoulli random sequence and $k(t)$ is the Gaussian distributed  noise process with zero mean and  $\sigma^{2}_{I}$ variance.
The received signal $\textbf{r}$ is expressed as
\begin{equation}\label{system1}
  \textbf{r}=\textbf{s}+\textbf{i}+\textbf{n} \in \mathbb{R}^{N},
\end{equation}
where $\textbf{s}$ is the TH-BPSK modulated desired multipath UWB signal, $\textbf{i}$ (discrete representation of $i(t)$ at Nyquist rate), and $\textbf{n}$ is  Gaussian (background) noise with zero mean, and
$\sigma^{2}_{n}$ variance.
The impulse noise, $\textbf{i}$, models impulse interference or harsh environment noise in the system and is sparse in nature \cite{san2016}. Hence, total effective noise power in the system can be written as $\sigma^{2}=\sigma^{2}_{n}+p \sigma^{2}_{I}$, where $p$ is the probability of impulse noise samples that occur in a given time duration and is expressed as $p=(\# \ \text{impulse noise samples})/N$. The signal-to-noise ratio (SNR) and signal-to-impulse noise ratio (SINR)  are defined as $\text{SNR}=\frac{\sigma^{2}_{s}}{\sigma^{2}_{n}}$ and $\text{SINR}=\frac{\sigma^{2}_{s}}{\sigma^{2}_{I}}$, respectively, where
$\sigma^{2}_{s}$ is the signal power and is considered unity, and $\sigma^{2}_{I} \gg \sigma^{2}_{n}$.  Further, in (\ref{system1}) inter-symbol-interference and inter-pulse-interference are assumed to be zero.

\vspace{-1em}
\section{Proposed receiver design}\label{sect3}

\subsection{Cluster detection algorithm}
This subsection presents a new cluster detection algorithm (CDA) for the proposed receiver design. It is known that the UWB signal cluster is symmetric around the maximum absolute peak value of the transmitted pulse \cite{molish2006, san2016, ncc2017}. This signal cluster symmetry can be used to differentiate  between  signal cluster and impulse noise samples. Since the symmetry of UWB signal is observed irrespective of the type of transmitted  pulse; the proposed method is independent of the type of UWB pulse and can be used for any UWB pulse.

Let $\mathcal{H}_{\textbf{i}}$ and $\mathcal{H}_{\textbf{s}}$ be the two hypothesis that label samples as impulse noise samples and desired signal samples frame by frame, respectively and are expressed as
\begin{equation} \label{hp1}
\begin{split}
\mathcal{H}_{\textbf{i}}: \textbf{r}= \textbf{s}+\textbf{i}+\textbf{n} \in \mathbb{R}^{N}, \\
 \mathcal{H}_{\textbf{s}}: \textbf{r}= \textbf{s}+\textbf{n} \in \mathbb{R}^{N}.
\end{split}
\end{equation}
 The maximum absolute peak value ($P^1_{max}$) and the corresponding time index ($I^{1}_{max}$) are calculated from the received signal $\textbf{r}$ and expressed as
 \begin{equation}\label{pr1}
 \left[P^1_{max}, I^{1}_{max}\right]=\max (|\textbf{r}|).
 \end{equation}
The sample $P^1_{max}=|\textbf{r}(I^{1}_{max})|$ belongs either to $\mathcal{H}_{\textbf{i}}$ or $\mathcal{H}_{\textbf{s}}$.
The classification of sample $\textbf{r}(I^{1}_{max})$ is done as
 \begin{equation}\label{pr2}
    |\textbf{r}(I^{1}_{max})-\textbf{r}(I^{1}_{max}+1)| \underset{\mathcal{H}_{\textbf{s}}}{\overset{\mathcal{H}_{\textbf{i}}}{\gtreqless}}  \mu,
\end{equation}
where $\mu$ is a constant that depends on the transmitted UWB pulse.
If the  sample $\textbf{r}(I^{1}_{max}) \in  \mathcal{H}_{\textbf{s}}$,  we conclude that no impulse noise is present in the signal $\textbf{r}$  and the peak value $P^1_{max}=|\textbf{r}(I^{1}_{max})| \ \in  \ \mathcal{H}_{\textbf{s}}$ represents the center of the first signal cluster detected at this position. In this case, we feed the signal $\textbf{r}$ to the conventional receiver for signal demodulation.  However, if $\textbf{r}(I^{1}_{max}) \notin  \mathcal{H}_{\textbf{s}}$, i.e., if $\textbf{r}(I^{1}_{max}) \in  \mathcal{H}_{\textbf{i}}$, then sample  $\textbf{r}(I^{1}_{max})$ represents the impulse noise sample and hence,
$\textbf{r}(I^{1}_{max})$ is assigned zero value to remove this impulse noise. Again, the maximum absolute peak value of the above modified signal $\textbf{r}$ (after assigning zero to impulse noise sample $\textbf{r}(I^{1}_{max})$) is calculated  and classified using  (\ref{pr1}) and (\ref{pr2}) respectively. This procedure is repeated until the $i^{\text{th}}$ maximum absolute peak valued sample $\textbf{r}(I^{i}_{max})$ of signal $\textbf{r}$ belongs to $\mathcal{H}_{\textbf{s}}$. Hence, a signal cluster is detected and the modified signal $\textbf{r}$ is applied to the conventional receiver for signal demodulation.
This CDA  is very simple and does not require multiplication or division operations. It requires only one subtraction per iteration for differentiating  between signal cluster and impulse noise samples and is summarized in \textbf{Algorithm} \textbf{\ref{algo}}.

The parameter $\mu$ in \textbf{Algorithm} \textbf{\ref{algo}} can be decided based on the transmitted UWB pulse $\textbf{w}$. Using the  maximum absolute peak value $P^{w}_{max}$  and the corresponding index $I^{w}_{max}$ of pulse $\textbf{w}$ at the transceiver, parameter $\mu$  can be selected such that
$\mu \geq |\textbf{w}(I^{w}_{max})-\textbf{w}(I^{w}_{max}-1)|$
(or $\mu \geq |\textbf{w}(I^{w}_{max})-\textbf{w}(I^{w}_{max}+1)|$ due to pulse symmetry).
The values of $\textbf{w}(I^{w}_{max})$ and $\textbf{w}(I^{w}_{max}\pm 1)$ are known apriori at the receiver in UWB communication system and an appropriately low value of $\mu$ can be selected according to the above expression for good system performance.
Further in \textbf{Algorithm} \textbf{\ref{algo}}, the  $\textbf{e}_{I^{i}_{max}} \in \mathbb{R}^{N}$ has entry `1' at $I^{i}_{max} $ position and `0's  at the remaining entries.

\vspace{-.5em}
\begin{algorithm}
\caption{Cluster-Detection Algorithm (CDA)}
\label{algo}
\begin{algorithmic}
\State
Initialize: $\mu \geq |\textbf{w}(I^{w}_{max})-\textbf{w}(I^{w}_{max}-1)|$, \ $i=1$\\
Input: received signal $\textbf{r} \in \mathbb{R}^{N}$ \\
Output: estimated signal $\hat{\textbf{s}} \in \mathbb{R}^{N}$ \\
Calculate: $[P^{i}_{max}, I^{i}_{max}]=\max (|\textbf{r}|)$  \\
\textbf{While}: $|\textbf{r}(I^{i}_{max})-\textbf{r}(I^{i}_{max}+1)|\geq \mu$\\
Update $\textbf{r}_{i}=\textbf{r}_{i}-\textbf{e}_{I^{i}_{max}}\textbf{r}_{i}$  \\
Set $i=i+1$\\
Calculate: $[P^{i}_{max}, I^{i}_{max}]=\max (|\textbf{r}|)$  \\
\textbf{End} \\
Update $\hat{\textbf{s}}=\textbf{r}$
\end{algorithmic}
\end{algorithm}

\vspace{-1em}
\subsection{False alarm and miss-detection probabilities}
The probability of false alarm $p_{f}$ can be calculated as
\begin{equation}\label{pf1}
p_{f} =   Pr\{|\textbf{r}(I^{i}_{max})-\textbf{r}(I^{i}_{max}+1)| \geq \mu |\mathcal{H}_{\textbf{s}}\}.
\end{equation}
Let $\tilde{r}_{s}|\mathcal{H}_{\textbf{s}}=\textbf{r}(I^{i}_{max})-\textbf{r}(I^{i}_{max}+1)=
\textbf{s}(I^{i}_{max})+\textbf{n}(I^{i}_{max})-\textbf{s}(I^{i}_{max}+1)-\textbf{i}(I^{i}_{max}+1)$
and $\tilde{r}_{s}|\mathcal{H}_{\textbf{s}}$ is distributed as  $\tilde{r}_{s}|\mathcal{H}_{\textbf{s}}\sim \mathcal{N}\left(0, 2((1-\rho_{s})\sigma_{s}^{2}+\sigma_{n}^{2})\right)$, where $\rho_{s}$ represents the correlation between two consecutive samples of signal $\textbf{s}$ while noise  samples are independent to each other.
The $p_{f}$ in (\ref{pf1}) can be written as
\begin{equation}\label{pf2}
p_{f} = \frac{1}{\sqrt{2\pi \sigma_{\tilde{r}_{s}}^{2}}} \int_{\mu}^{\infty} \exp^{-\frac{x^2}{2\sigma_{\tilde{r}_{s}}^{2}}}dx+\frac{1}{\sqrt{2\pi \sigma_{\tilde{r}_{s}}^{2}}} \int_{-\infty}^{-\mu} \exp^{-\frac{x^2}{2\sigma_{\tilde{r}_{s}}^{2}}}dx,
\end{equation}
where $\sigma_{\tilde{r}_{s}}^{2}=2((1-\rho_{s})\sigma_{s}^{2}+\sigma_{n}^{2})$.
Therefore, $p_{f}=2 Q{\left(\frac{\mu}{\sqrt{2((1-\rho_{s})\sigma_{s}^{2}+\sigma_{n}^{2})}}\right)}$.
Similarly, the probability of miss-detection $p_{m}$ is expressed as
\begin{equation}\label{pf3}
p_{m} =    Pr\{|\textbf{r}(I^{i}_{max})-\textbf{r}(I^{i}_{max}+1)| < \mu |\mathcal{H}_{\textbf{i}}\}.
\end{equation}
Let $\tilde{r}_{i}|\mathcal{H}_{\textbf{i}}=\textbf{r}(I^{i}_{max})-\textbf{r}(I^{i}_{max}+1)$ and $\tilde{r}_{i}|\mathcal{H}_{\textbf{i}}$ is distributed as  $\tilde{r}_{i}|\mathcal{H}_{\textbf{i}}\sim \mathcal{N}(0, 2((1-\rho_{s})\sigma_{s}^{2}+\sigma_{n}^{2}+p\sigma_{I}^{2}))$. After some intermediate steps, $p_{m}$ can be written as
$p_{m}=1-2 Q{\left(\frac{\mu}{\sqrt{2((1-\rho_{s})\sigma_{s}^{2}+\sigma_{n}^{2}+p\sigma_{I}^{2})}}\right)}$.
The proposed impulse noise rejection method select the parameter $\mu$ based on the transmitted UWB pulse. Hence, the proposed method does not need to find the optimal threshold, unlike clipper or blanking based receiver.
In general, optimal threshold using received signal statistics, such as signal and noise power, is difficult to compute.

\subsection{Convergence analysis of the proposed CDA}
Let $\textbf{s}, \textbf{r} \in \mathbb{R}^{N}$ be the desired UWB and received   signals in the frame for a particular data symbol.
In the proposed CDA, signal for  $(i+1)^{\text{th}}$ iteration is written as
$\textbf{r}_{i+1}= \textbf{r}_{i}- \textbf{e}_{I^{i}_{max}}\textbf{r}_{i}, i=1,2,... $ where $\textbf{e}_{I^{i}_{max}} \in \mathbb{R}^{N}$ has entry `1' at the $I^{i}_{max}$ position and `0's at the remaining entries. Further, $\lVert \textbf{r}_{i+1}\rVert_{2}^{2}=\lVert \textbf{r}_{i}- \textbf{e}_{I^{i}_{max}}\textbf{r}_{i}\rVert_{2}^{2}=\lVert \textbf{r}_{i}\rVert_{2}^{2}- \textbf{r}_{i}(I^{i}_{max})^{2}$.
Therefore, $\lVert \textbf{r}_{i+1}\rVert_{2} <\lVert \textbf{r}_{i}\rVert_{2}$ (where $\textbf{r}_{i}(I^{i}_{max})^{2}\neq 0$) and can also be written as
$\lVert \textbf{r}_{i+1}-\textbf{s}\rVert_{2} =\beta\lVert \textbf{r}_{i}-\textbf{s}\rVert_{2}$, where $\beta \in (0,1)$.
The distance between  signal $\textbf{r}_{i}$  and desired signal $\textbf{s}$ in the $i^{\text{th}}$ iteration is written as $ \lVert \textbf{r}_{i} -\textbf{s} \rVert_{2}= \beta^{i} \lVert \textbf{r} -\textbf{s}  \rVert_{2}$. Hence, as $i \rightarrow \infty$, $\lVert \textbf{r}_{i} -\textbf{s} \rVert_{2} \rightarrow 0$ i.e. $\textbf{r}_{i} \rightarrow \textbf{s}$.

In  practical implementation, the proposed algorithm will have some finite  distance  between the desired signal $\textbf{s}$  and the received signal $\textbf{r}_{i}$ after $i^{\text{th}}$ iteration and hence, can be expressed as $\lVert \textbf{r}_{i} -\textbf{s} \rVert_{2} \rightarrow \epsilon_{0}$, where $\epsilon_{0}\geq 0$. The parameter $\epsilon_{0}$ depends on the SNR, number of iterations, and parameter $\mu$.

\subsection{BER performance}
In this subsection, we have analyzed the BER performance of the proposed receiver.
Let  $\hat{\textbf{s}}$ be the output of the CDA. The  signal $\hat{\textbf{s}}$  includes the background Gaussian noise and hence, $\left\Vert\hat{\textbf{s}}-\textbf{s}\right\Vert_{2} \geq 0$.
Therefore,
signal $\hat{\textbf{s}}$ can be  written as   $\hat{\textbf{s}}=\textbf{s}+\textbf{e}$, where $\textbf{e}$ is the undesired (noise) additive Gaussian noise in the signal $\hat{\textbf{s}}$.
The pdf of $\textbf{e}$ is Gaussian distributed and is given by $\mathcal{N}(0, \sigma^{2}_{e})$. In general, $\sigma^{2}_{e} \geq \sigma^{2}_{n}$ because a few samples of impulse noise may appear similar in amplitude to Gaussian background noise and hence, may not have been filtered out by \textbf{Algorithm} \textbf{\ref{algo}} and still be present in the output signal $\hat{\textbf{s}}$. The probability  of overlap of the desired signal and impulse noise samples is  low due to the sparse nature of both $\textbf{s}$ and $\textbf{i}$. Therefore, assigning zero value to the desired signal sample during cluster detection is almost zero. This will mostly not lead to any desired signal power deterioration in
the proposed receiver design. However,  the signal power deterioration in case of  UWB signal blanking due to overlapping with impulse noise is analyzed in the next subsection.

This paper considers correlation-based coherent receiver for data symbol detection. Thus, the correlator output $\zeta$ for a positive transmitted data symbol  is written  as
\begin{equation}\label{cr1}
\zeta=\langle\textbf{s}+\textbf{e}, \boldsymbol \phi \rangle,
\end{equation}
where $\boldsymbol \phi$ is the template signal. The template signal is generated using UWB pulse $\textbf{w}$ and channel impulse response (CIR) $\textbf{h}$ with known time hopping code as $\boldsymbol \phi=\textbf{h}\ast \textbf{w}$.
Correlator output  $\zeta$  is  Gaussian distributed, i.e.,
\begin{equation}\label{cr2}
\zeta \sim \mathcal{N}( \lVert\textbf{w}\rVert^{2}_{2} \sum_{l=0}^{L-1}|\alpha_{l}|^{2}, \lVert\textbf{w}\rVert^{2}_{2}\sigma^{2}_{e}\sum_{l=0}^{L-1}|\alpha_{l}|^{2}),
\end{equation}
where $\alpha_{l}$ is the channel coefficient of $l^{\text{th}}$ path and $L$ is the total number of resolved paths in CIR $\textbf{h}$.
The bit error  probability $p_{pr}(e|\textbf{h})$ in the presence of impulse noise for the given CIR $\textbf{h}$ using the proposed correlator based receiver design in TH-BPSK system is given as
\begin{equation}\label{ber_pr}
p_{pr}(e|\textbf{h})=Q\left( \sqrt{\frac{ (1-\rho)\lVert\textbf{w}\rVert^{2}_{2}\sum_{l=0}^{L-1}|\alpha_{l}|^{2}}{\sigma^{2}_{e}}} \right),
\end{equation}
where $Q(\cdot)$ is the tail probability of normal Gaussian distribution and all the transmitted symbols are equally likely in (\ref{ber_pr}).
In the absence of impulse noise ($\sigma^{2}_{e}=\sigma^{2}_{n}$), $ p_{pr}(e|\textbf{h})$ in (\ref{ber_pr}) corresponds to the conventional TH-BPSK system.  For the AWGN channel, (\ref{ber_pr}) is expressed as
$ p_{pr}(e)= Q\left( \sqrt{\frac{(1-\rho)\lVert\textbf{w}\rVert^{2}_{2}}{\sigma^{2}_{e}}} \right)$.
The factor $\rho$ depends on the blanking of  UWB signal samples and $\rho\rightarrow 0$ as the sparsity of UWB signal $\textbf{s}$  and/or multipath channel diversity increases
for a fixed sparsity level of impulse noise.

\subsection{UWB signal and impulse noise samples overlap}
In this subsection, the  effect of overlapping impulse noise sample on UWB signal is analyzed for the proposed CDA based receiver.
The number of samples, $\Omega$, in a frame duration, $T_{f}$, at the sampling frequency, $F_{s}$, can be expressed as $\Omega=\lceil T_{f} \times F_{s}\rceil$, where $\lceil (\cdot)\rceil$ represents a ceiling of $(\cdot)$. The total number of samples of desired UWB signal and impulse noise  in a frame duration is written as $\Omega_{\textbf{s}}=\lceil L \Omega_{\textbf{w}} \rceil$ and $\Omega_{\textbf{i}}=\lceil p \Omega \rceil$, respectively, where $\Omega_{\textbf{w}}$  is the non-zero samples in UWB pulse $\textbf{w}$.
Due to the sparse nature of UWB signal and impulse noise,
$\Omega >>\Omega_{\textbf{s}} >>\Omega_{\textbf{i}}$ and their occupancy rate in a frame  is expressed as $\Omega_{\textbf{s}}/\Omega$ and $\Omega_{\textbf{i}}/\Omega$, respectively.
The probability of  a single impulse noise sample's chance of occurrence  in the desired UWB signal cluster's duration is written as
$p_{\textbf{s},\textbf{i}}=\tilde{\Omega}_{\textbf{i}}/\Omega_{\textbf{w}}$, where $\tilde{\Omega}_{\textbf{i}}=\Omega_{\textbf{i}}/L, \ \tilde{\Omega}_{\textbf{i}}< \Omega_{\textbf{w}}$ and $\tilde{\Omega}_{\textbf{i}}$ relative impulse noise samples occupancy in a single UWB signal cluster.
Therefore, the probability that $k$-number of clusters have impulse noise is expressed as
$p_{\textbf{s},\textbf{i},k}=\sum_{k=1}^{L}\binom{L}{k} p_{\textbf{s},\textbf{i}}^{k}(1-p_{\textbf{s},\textbf{i}})^{L-k}$, where $\binom{L}{k}$ is a binomial coefficient.
Thus, probability $p_{\textbf{s},\textbf{i},k}$ (all clusters have impulse noise) reduces as $p$ decreases or $L$ increases.
Hence, the desired UWB signal sample's blanking probability $p_{\textbf{s},\textbf{i},k}$  is small for a multipath channel as compared to an AWGN channel.
Further, the desired UWB signal energy loss  in a cluster $E_{\textbf{s},loss,l}, l=1,2,...,L$ due to blanking of a signal sample in the receiver design is expressed in the range of $ \lVert\alpha_{min}\textbf{w}_{min}  \lVert^{2}_{2}$ to $ \lVert \alpha_{max}\textbf{w}_{max} \lVert^{2}_{2}$, where $\alpha_{min}=\min_{l} \{\alpha_{l}\}_{l=1}^{L}, \  \alpha_{max}=\max_{l} \{\alpha_{l}\}_{l=1}^{L},\
\textbf{w}_{min}= \min _{i} \{\textbf{w}_{i}\}_{i=1}^{\Omega_{\textbf{w}}}$ and
$\textbf{w}_{max}= \max _{i} \{\textbf{w}_{i}\}_{i=1}^{\Omega_{\textbf{w}}}$.
Therefore, effective signal energy  loss in a frame is expressed as $E_{\textbf{s},loss}=p_{\textbf{s},\textbf{i},k} E_{\textbf{s},loss,l}$ and is smaller  for a multipath channel as compared to an AWGN channel, i.e., $E_{\textbf{s},loss, \text{multipath}}\leq E_{\textbf{s},loss, \text{AWGN}}$ due to the low value of $p_{\textbf{s},\textbf{i},k}$.
In other words, received signal power spread in more number of low energy pulses
in multipath channel, hence, the  blanking of a single sample results in  very less signal energy loss as compared AWGN channel, which has more energy concentration in the  single received pulse.
Hence, in this work multipath channel diversity (which reduces the effective value of  $p_{\textbf{s},\textbf{i},k}$) and the desired received UWB signal sparsity (sparsity reduces the overlapping probability of the signal and impulse noise)
add  robustness against the blanking loss in the proposed UWB receiver design.

\vspace{-1.2em}
\section{Simulation Results and Discussion}\label{sect4}
This section presents performance evaluation of the proposed receiver design compared to the conventional and the non-linear blanking based receivers. In \cite{zhidkov2006, juwono2016performance,epple2016advanced}, non-linear blanking receiver is analyzed for mitigating impulse noise effect in OFDM system. However, we use it for UWB system with suitable modification and parameter selection for the first time in UWB literature in order to mitigate impulse noise. All the simulations are carried out for the TH-BPSK UWB system for $F_{s}=16$ GHz sampling frequency using the second derivative Gaussian pulse $\textbf{w}$ \cite{san2016,sharmanew,ding2013first} with the pulse width parameter $\tau=0.4$ nanoseconds and with single frame per data symbol. The transmitter and receiver synchronization is assumed with the perfect channel state information  at the receiver.
The TH code is generated using chip duration of $1$
nanoseconds and cardinality of $3$ for
AWGN and $6$ for multipath IEEE 802.15.4a channels.
In simulation results, legend ``BPSK" represents the BER performance of the conventional receiver in the impulse noise free system, ``Theory" represents semi-analytical results using
(\ref{ber_pr})  and, ``BR" and ``CDA" represent  BER performance using the blanking receiver \cite{zhidkov2006, juwono2016performance,epple2016advanced} and  the proposed CDA receiver in the presence of impulse noise, respectively.

\begin{figure}[h]
\vspace{-1em}
\centering{
\includegraphics[height=100mm,width=85mm]{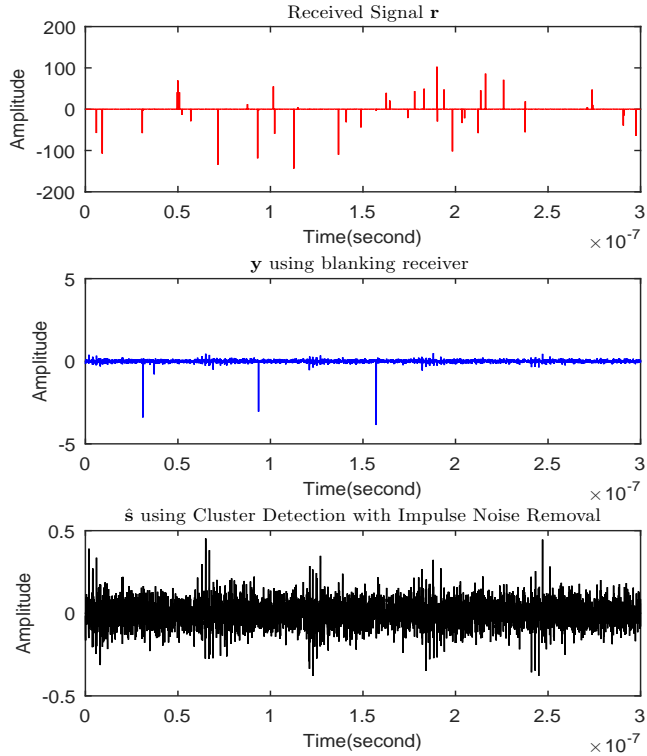}}
\caption{\small The received signal $\textbf{r}$ (at the top), blanking output signal $\textbf{y}$ (using \cite{zhidkov2006, juwono2016performance,epple2016advanced}) and signal $\hat{\textbf{s}}$ (at the bottom) at the output of the proposed CDA. }
\label{signal}
\vspace{-1em}
\end{figure}

The received signal $\textbf{r}$, blanking output signal $\textbf{y}$ (using the blanking method in \cite{zhidkov2006, juwono2016performance,epple2016advanced, rabie2014}), and the output of the proposed CDA algorithm $\hat{\textbf{s}}$ for five frame time duration in multipath channel model CM1 \cite{molish2006} are shown in Fig. \ref{signal}. The blanking non-linearity is applied to the received signal  $\textbf{r}$. Samples of $\textbf{r}$ are assigned zero value if $|\textbf{r}_{i}| \geq T, \ i=1,2,...,N$, where $T$ is a constant threshold value.
To mitigate impulse noise effect, threshold $T$ for the blanking based receiver in UWB system is selected such that false alarm and miss-detection probabilities are minimized. The optimal value of  $T$ is derived as \cite{zhidkov2006, juwono2016performance,epple2016advanced}
\begin{equation}\label{th1}
T_{opt}=\min_{T}  \left\{ Pr(\mathcal{H}_{\textbf{s}}) p_{f,T}+
Pr(\mathcal{H}_{\textbf{i}}) p_{m,T} \right\}.
\end{equation}
Similar to $p_{f}$ (in eq (\ref{pf1})) and $p_{m}$ (in eq (\ref{pf3})), $p_{f,T}$  and $p_{m,T}$ are calculated and expressed as $2 Q{\left(\frac{T}{\sqrt{\sigma_{s}^{2}+\sigma_{n}^{2}}}\right)}$
and
$\left(1-2Q{\left(\frac{T}{\sqrt{\sigma_{s}^{2}+\sigma_{n}^{2}+p\sigma_{I}^{2})}}\right)}\right)$,
respectively.
The exact solution  of (\ref{th1})  is difficult due to multiple $Q(\cdot)$ functions. Hence, the $Q(\cdot)$ function is approximated
using a method in \cite{karagiannidis}. On equating the derivative of (\ref{th1}) to zero, a sub-optimal value of $T$ is obtained. For example, at an $\text{SINR}=-40$ dB with $p=0.01$, sub-optimal values of $T_{opt}$ equal to 4 and 2.5 are obtained for
$\text{SNR}$ of -2 and 5 dB,  respectively.
In simulations, we have used fixed value of $T$ throughout the entire range of SNR. However,
SNR specific $T$ can be selected using a look-up table method at the receiver, which requires frame based SNR estimation, thereby  increasing computational complexity of the receiver.

In Fig. \ref{signal}, we have considered $\text{SINR}=-40$ dB, $\text{SNR}=20$ dB, blanking threshold $T=4$ (for blanking receiver), and impulse noise probability $p=0.01$ in this simulation setup. The amplitude of impulse noise samples is very high as observed in Fig. \ref{signal} (top subfigure) and the desired signal $\textbf{s}$ is completely buried within the impulse noise. In the blanking based receiver \cite{zhidkov2006, juwono2016performance,epple2016advanced}, high amplitude samples  of  impulse noise are blanked (assigned zero value), while low amplitude impulse samples are present at the output of  blanking unit in signal as shown in Fig. \ref{signal} (middle subfigure).
Hence,  performance of the blanking based receiver deteriorates due to the  presence of few impulse noise samples and is sensitive to the threshold value $T$.
On the other hand, all the samples of impulse noise  are removed with the proposed algorithm without any modification in the desired signal as observed in Fig. \ref{signal} (bottom subfigure).

\begin{figure}[h]
\vspace{-.5em}
\centering{
\includegraphics[height=75mm,width=125mm, trim = 25 0 0 0]{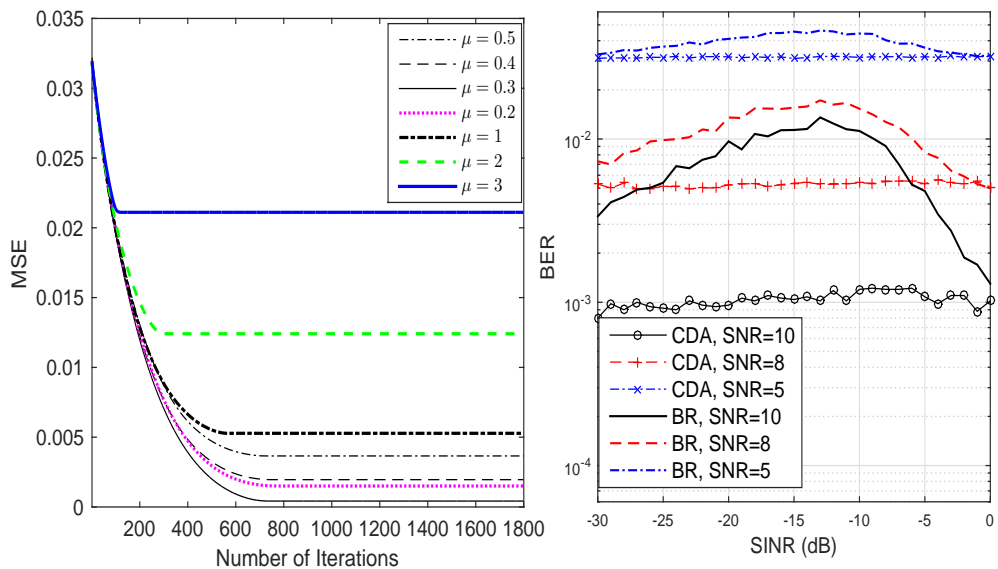}}
\captionsetup{justification=centering}
\caption{\footnotesize{ Performance analysis of the proposed receiver}}
\label{mse2}
\vspace{-1em}
\end{figure}

Further, the mean square error (MSE) between desired multipath received signal $\textbf{s}$ and CDA output signal $\hat{\textbf{s}}$ is calculated and defined as
\begin{equation}\label{error11}
\text{\textit{MSE}}=\frac{\lVert\textbf{s} -\hat{\textbf{s}} \rVert_{2}^{2}}{\tilde{N}},
\end{equation}
where $\tilde{N}$ is the number of samples in $\textbf{s}$.
In (\ref{error11}), signal $\textbf{s}$ is fixed and CDA output signal $\hat{\textbf{s}}$ changes after each iteration. Hence, MSE in (\ref{error11}) changes after each iteration. Simulation results are plotted in Fig. \ref{mse2} (left) for $\text{SINR}=-40$ dB and $\text{SNR}=20$ dB  in multipath communication channel model CM1 using various values of the parameter $\mu$ in \textbf{Algorithm} \textbf{\ref{algo}}.
The rate of decrease in MSE with the number of iteration is same for all the values of $\mu$. However, high values of $\mu=1, 2, 3$ saturate at higher error floor as compared to lower values ($\mu=0.2, 0.3, 0.4, 0.5$) as observed from Fig. \ref{mse2} (left). Further, $\mu=0.3$ provides the lowest value of error floor as shown in Fig. \ref{mse2} (left).
Based on empirical results, $\mu=\kappa |\textbf{w}(I^{w}_{max})-\textbf{w}(I^{w}_{max}-1)|$, where  $\kappa \in [2 \ \ 3]$.
 Further, at a constant threshold value, BER performance of the blanking based receiver (\cite{zhidkov2006, juwono2016performance,epple2016advanced}) varies with SINR as observed in Fig. \ref{mse2} (right). On the other hand, the proposed receiver provides optimum results irrespective of the values of SINR chosen as observed in
Fig. \ref{mse2} (right).
Further, BER performance of both the proposed and blanking receivers converge around $\text{SINR}=0$ dB due to
low amplitude of impulse noise samples at these SINR values and hence, impulse noise behaves similar to Gaussian noise at $\text{SINR}=0$ dB.

\begin{figure}[h]
\vspace{-.7em}
\centering{
\includegraphics[height=65mm,width=100mm]{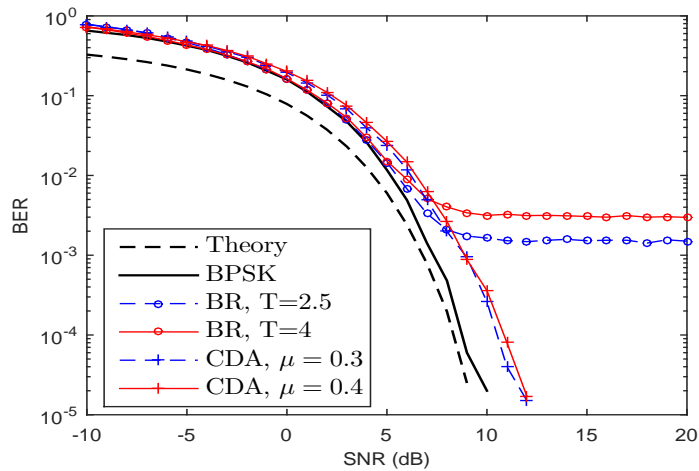}}
\caption{ Average BER vs. SNR performance of TH-BPSK system using the proposed and the blanking based receiver (\cite{zhidkov2006, juwono2016performance,epple2016advanced}) in the presence of impulse noise in AWGN channel.}
\label{awgn+cm1}
\end{figure}

\begin{figure}[h]
\centering{
\includegraphics[height=65mm,width=100mm]{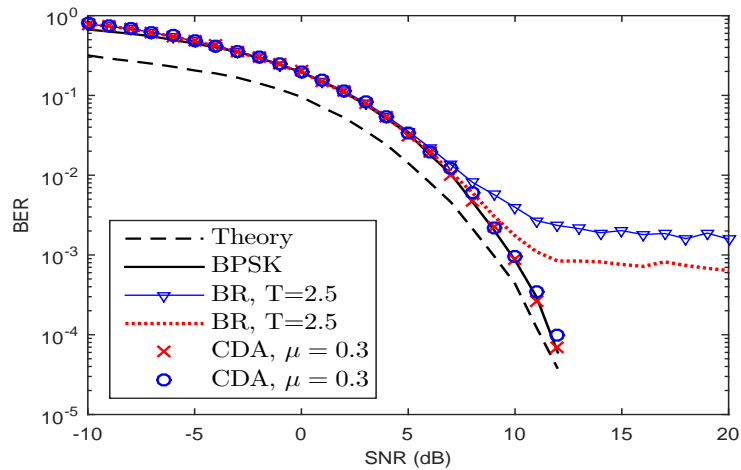}}
\caption{Average BER vs. SNR performance of TH-BPSK system using the proposed and the blanking based receiver (\cite{zhidkov2006, juwono2016performance,epple2016advanced}) in the presence of impulse noise in  CM1 (blue and red lines correspond to $\text{SINR}=-20, -30$ dB, respectively) channels.}
\label{11}
\end{figure}

Next, the BER performance of TH-BPSK UWB system in the presence of impulse noise using the proposed receiver and the blanking non-linearity based receiver in \cite{zhidkov2006, juwono2016performance,epple2016advanced}   is analyzed in AWGN and multipath IEEE 802.15.4a channel CM1 \cite{molish2006}. Results are shown in Fig. \ref{awgn+cm1} and Fig. \ref{11}.
In AWGN channel, $\text{SINR}=-30$ dB, $p=0.01$, $T=2.5,4$ (for blanking receiver), and frame duration $T_{f}=10$ nanoseconds are chosen, while in CM1 channel, $\text{SINR}=-20, -30$ dB, $p=0.01$, $T=2.5$, and  $T_{f}=60$ nanoseconds are considered.
The blanking receiver exhibits bit error floor for both the values of threshold and SINR in the presence of impulse noise in both AWGN and CM1 channels as shown in Fig. \ref{awgn+cm1} and and Fig. \ref{11}.
The BER performance of the proposed receiver in the presence of impulse noise is close to the BER performance of the conventional receiver (BPSK) in impulse noise free scenario and is free from any bit error floor as observed in Fig. \ref{awgn+cm1} and Fig. \ref{11} respectively.
Further, CDA based receiver's BER performance is degraded marginally in AWGN channel due to non-zero (but small)
$p_{\textbf{s},\textbf{i},k}$ unlike in the multipath channel, which has $p_{\textbf{s},\textbf{i},k}$ is close to zero.
Proposed receiver needs around $\lceil p \Omega \rceil$ iterations per frame, hence it is computationally efficient and free from any SNR dependent threshold value selection like blanking receivers.

\section{Conclusion}\label{sect5}
A signal cluster sparsity based receiver design  for the impulse noise mitigation in a UWB system is proposed.
 The proposed  receiver is observed to be robust and has improved  bit error rate  performance (close to the impulse noise free system)  as compared to the blanking non-linearity based receiver  in the presence of Bernoulli-Gaussian impulse noise for single and multipath channels. The work presented in this paper is helpful for robust operation and  analysis of UWB based devices  such as WSNs, IoTs, and M2M that work extensively in harsh impulse noise environments and hence, require robust receiver designs in practical applications.
In future, the proposed cluster sparsity based receiver design can be extended for multiuser communication in the presence of impulse noise environment.

\bibliographystyle{ieeetran}
\footnotesize
\bibliography{references_letter}

\end{document}